\documentstyle[twocolumn,floats,aps,prl,epsfig]{revtex}
\tighten
\begin{document}
\draft
\title{Small, Dense Quark Stars from Perturbative QCD}
\author{Eduardo S. Fraga$^a$, Robert D. Pisarski$^a$, and J\"urgen
  Schaffner-Bielich$^b$} 
\address{$^a$Department of Physics, Brookhaven National Laboratory, 
Upton, NY 11973-5000, USA \\
$^b$RIKEN BNL Research Center, Brookhaven National Laboratory, 
Upton, NY 11973-5000, USA}
\date{\today}
\maketitle   


\begin{abstract}
As a model for nonideal behavior in the equation of state
of QCD at high density, we consider cold quark matter in
perturbation theory.  To second order in
the strong coupling constant, $\alpha_s$, 
the results depend sensitively on the choice of the renormalization
mass scale.  Certain choices of this scale correspond
to a strongly first order chiral transition,
and generate quark stars with maximum masses and radii
approximately half that of ordinary neutron stars.
At the center of these stars, quarks are essentially massless.
\end{abstract}



Strongly interacting matter under extreme conditions can reveal new 
phenomena in Quantum Chromodynamics (QCD).
Compact stars serve as an excellent observatory to probe QCD at
large density, as their interior might be dense enough to allow
for the presence of chirally symmetric quark matter, {\it i.e.}, quark stars 
\cite{G_b,Free,Baluni,bod,wit,bag,thirda,othera,otherb,thirdb,njl}.

The usual model used for quark stars is
a bag model, with at most a correction $\sim \alpha_s$ from
perturbative QCD \cite{bag}.
In the massless case, the first order correction cancels out in the
equation of state, so that one ends up finally with a free gas of quarks 
modified only by a bag constant.  If the bag constant is fit from
hadronic phenomenology, then the gross features of quark stars are very
similar to those expected for neutron stars:
the maximum mass is $\approx 2. M_\odot$, with a radius $\approx 10$~km.  

In this article we consider quark stars, using the equation of
state for cold, dense QCD in perturbation theory 
to $\sim \alpha_s^2$ \cite{Free,Baluni}.  These results are well
known, and our only contribution is to use modern determinations
of the running of the QCD coupling constant \cite{PDG}.  
At the outset, we stress that we do {\it not} suggest that the
perturbative equation of state is a good approximation for
the densities of interest in quark stars.  Rather, we use it merely
as a model for the equation of state of QCD.

To $\sim \alpha_s^2$, there is significant sensitivity to the choice of
the renormalization mass scale.  Under our assumptions,
we find that this choice is tightly constrained by the physics.
We consider two illustrative values of this parameter.  One
choice corresponds to a weakly first order chiral transition (or
no true phase transition),
and gives maximum masses and radii very similar to that of 
neutron stars.  
The second choice corresponds to a strongly
first order chiral transition \cite{strong}, and generates two types of stars.
One type has densities a few times that of nuclear matter,
and looks like the stars of a weakly first order chiral
transition.  In addition, however, there is a new 
class of star \cite{thirda,thirdb},
with densities much higher than that of nuclear matter .  For this
new class, the 
maximum mass is $\approx 1. M_\odot$, with a radius $\approx 5$~km.  
Other models with nonideal behavior also generate small, dense
quark stars \cite{othera,otherb}.

Assume that the chiral phase transition occurs at a chemical
potential $\mu_\chi$ \cite{phase}.    
Our perturbative equation of state is applicable 
only in the chirally symmetric phase, when the quark
chemical potential $\mu > \mu_\chi$.
In this phase, the effects of a strange quark mass,
$m_s\approx 100$~MeV \cite{Blum99}, are small relative to the
quark chemical potentials, $\mu > 300$~MeV.  Thus
we take three flavors of massless quarks with equal
chemical potentials \cite{G_b,raj}.

The thermodynamic potential of a plasma of massless quarks 
and gluons was calculated perturbatively to $\sim \alpha_s^2$
by Freedman and McLerran \cite{Free} and by Baluni \cite{Baluni},
using the momentum-space subtraction (MOM) scheme.
The MOM coupling is related to that in the
modified mimimal substraction scheme, $\overline{{\rm MS}}$, as
\cite{Baluni,explanation,Baier,Blaizot_1}:
\begin{equation}
\frac{\alpha_s^{{\rm MOM}}}{\pi}=\frac{\alpha_s^{\overline{{\rm MS}}}}{\pi}\;
\left[ 1+{\cal A}~\frac{\alpha_s^{\overline{{\rm MS}}}}{\pi}\right] \quad ;
\label{e1}
\end{equation}
$\alpha_s = g^2/(4 \pi)$, with $g$ the QCD coupling constant,
and ${\cal A}=151/48 - (5/18) N_f$, with $N_f$ the number
of massless flavors.  (While we take $N_f = 3$, 
we give formulas for arbitary $N_f$.)
In the $\overline{{\rm MS}}$ scheme, to $\sim \alpha_s^2$ the
thermodynamic potential is then
\begin{eqnarray}
\label{e2}
\lefteqn{\Omega(\mu)= - \frac{N_f \mu^4}{4\pi^2}
\left\{1-2 \left(\frac{\alpha_s}{\pi}\right) \right.} \nonumber \\
&-& 
\left. \left[G+N_f\ln{\frac{\alpha_s}{\pi}}
+ \left(11-\frac{2}{3} N_f \right) \ln{\frac{\bar\Lambda}{\mu}} \right]
\left(\frac{\alpha_s}{\pi}\right)^2 \right\} \; ,
\label{eq:omega}
\end{eqnarray}
where $G=G_0-0.536N_f+ N_f\ln{N_f}$, $G_0=10.374 \pm .13$
\cite{error}, and
$\bar\Lambda$ is the renormalization subtraction point.  
In $\overline{{\rm MS}}$ scheme, the thermodynamic potential is
manifestly gauge invariant.
We take the scale dependence of the strong coupling constant,
$\alpha_s \equiv \alpha_s(\bar\Lambda)$ as \cite{PDG,qual}:
\begin{eqnarray}
\label{e3}
\alpha_s(\bar\Lambda)&=&\frac{4\pi}{\beta_0 u}
\left[ 1 - \frac{2\beta_1}{\beta_0^2}
\frac{\ln (u)}{u}
\right.\nonumber \\
&+&  \frac{4\beta_1^2}{\beta_0^4 u^2}
\left.\left( \left(\ln(u)-\frac{1}{2}\right)^2 +
\frac{\beta_2 \beta_0}{8\beta_1^2} - \frac{5}{4} \right) 
\right] \quad ;
\end{eqnarray}
$u = \ln(\bar\Lambda^2/\Lambda^2_{\overline{{\rm MS}}})$, 
$\beta_0=11-2 N_f/3$, $\beta_1=51-19 N_f/3$, and
$\beta_2=2857 - 5033 N_f/9 + 325 N_f^2/27$.
The scale $\Lambda_{\overline{{\rm MS}}}$ is fixed by 
requiring that $\alpha_s=0.3089$ at
$\bar\Lambda=2$ GeV \cite{PDG}; for $N_f = 3$,
(\ref{e3}) gives $\Lambda_{\overline{{\rm MS}}}=365$~MeV.

All thermodynamic quantities follow consistently
from $\Omega(\mu)$.  The pressure is given by $p(\mu)=-\Omega(\mu)$,  
the quark number density by
$n(\mu) = (\partial p / \partial\mu)$, and the energy density by
$\epsilon = -p + \mu n$.  Given our stated assumptions, the only
freedom we have in the model is the choice of the ratio $\bar\Lambda/\mu$.
To illustrate this, we take the values $\bar\Lambda/\mu = 1,2,3$.

For reasons which will become clear later, 
we find the choice $\bar\Lambda = 2 \mu$ especially interesting.
Start with a very large chemical potential, such as 
$\mu = 100$~GeV, for which $\alpha_s \sim .095$
(for the purposes of discussion, assume $N_f=3$ at this scale).  
The first order term decreases the ideal gas pressure by $\sim 6\%$;
the sum of the first and second order terms decrease the
pressure by $\sim 7\%$ of the ideal gas value.
Because the strong coupling constant runs relatively slowly
with $\mu$ at large $\mu$, even at $\mu = 1$~GeV, where
$\alpha_s \sim .31$, the
first order term decreases the ideal gas pressure only by 
$\sim 20\%$; the first and second order terms, by $\sim 30\%$.  

As can be seen from (\ref{e2}), 
the perturbative expansion of the thermodynamic
potential is an expansion in a power series not just in
$\alpha_s$, but in $\alpha_s \log(\alpha_s)$.
The logarithm of $\alpha_s$ arises from the plasmon effect, where
the Debye mass squared $m_{D}^2 \sim \alpha_s \mu^2$.
Because gluons at $T=0$ have four-dimensional phase space
in loop integrals, however, the plasmon effect is relatively innocuous,
and only produces logarithms, $\log(m_{D}/\mu) \sim \log(\alpha_s)$.

This is in stark contrast to the perturbative expansion of the
free energy at nonzero temperature, $T \neq 0$.
While there is again a plasmon effect, 
$m_{D}^2 \sim \alpha_s T^2$, because in loop integrals
static gluons at $T \neq 0$
have a three dimensional phase space,
the perturbative expansion is not in $\alpha_s$,
but in $\sqrt{\alpha_s}$.
The series in $\sqrt{\alpha_s}$ is much worse behaved than that
at $\mu \neq 0$, $T = 0$, and does not converge until very
high temperatures \cite{hot_pert}.  
The convergence appears to
improve after resummation \cite{resum,Baier,Blaizot_1},
or by using Pad\'e methods \cite{pade}.

Consequently, the perturbative series for the thermodynamic potential
may be much better behaved at $\mu \neq 0$ (and $T=0$) than 
at $T \neq 0$ \cite{Rob00}.  This does not imply that a given
value of $\alpha_s$, which is adequate to compute the thermodynamic
potential, works equally well for all other quantities.  In particular,
the gaps for color superconductivity are nonperturbative,
$\phi \sim \exp(-1/\sqrt{\alpha_s})$ \cite{son,superreview},
and much smaller values of
$\alpha_s$ appear to be required to reliably compute them
\cite{Rajagopal00}.  
In QCD, effective models find that even when
$\mu \sim 400$~MeV, these gaps are at most 
$\sim 100$~MeV \cite{superreview}.
As the relative change in the thermodynamic potential is only
$\sim (\phi/\mu)^2$, then, 
for the equation of state in QCD, color superconductivity 
is never a large effect.

To truly know how well perturbation theory converges at $\mu \neq 0$,
it is imperative to compute the thermodynamic potential
to $\sim \alpha_s^3$.  Unlike the case of $T \neq 0$, which
is sensitive to nonperturbative effects from static magnetic gluons
from $\sim \alpha^3_s$ on, at $\mu \neq 0$ (and $T=0$), the entire
power series in $\alpha_s$ is well defined \cite{son}.

\begin{figure}[t]
\centerline{\epsfig{figure=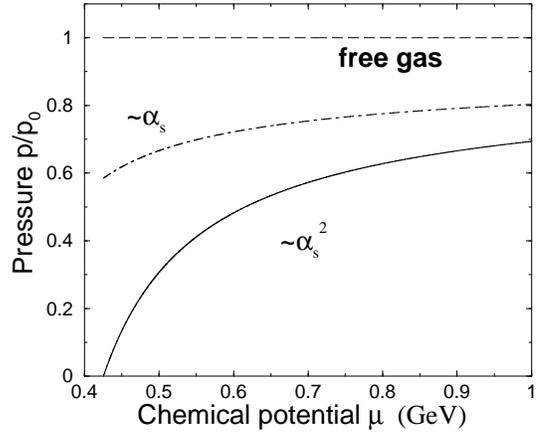,width=7cm}}
\caption{The total pressure, relative to the pressure of an
ideal gas, $p_0$; including terms to order
$\sim \alpha_s$ and to order $\sim \alpha_s^2$, as a function 
of $\mu$; $\bar\Lambda=2\mu$.}
\label{fig:pressure}
\end{figure}

We now use the perturbative calculation of the thermodynamic potential
for $\mu < 1$~GeV.  
Since both terms $\sim \alpha_s$ and $\sim \alpha_s^2$ have negative
coefficients, as $\alpha_s(\mu)$ increases with decreasing $\mu$,
eventually the pressure vanishes.  While it is clearly invalid using
perturbation theory when $p = 0$, it at least provides
a well defined model of dense QCD.  
In Fig.~\ref{fig:pressure} we show the pressure for
$\bar\Lambda=2\mu$; it vanishes at $\mu_c = 425$~MeV,
where $\alpha_s \sim .65$.  This corresponds
to a quark density $\sim 4.35 \rho_0$, where $\rho_0$ is the 
density of quarks in nuclear matter, $\sim 3 \times .16$/fm$^3$.

A weakness in our model is how to match the equation of state
for massless quarks, (\ref{e2}), 
onto that for massive quarks and hadrons.  
The quark chemical potential must be larger than one third
of the nucleon rest mass, minus one third the binding energy
of nuclear matter, $\mu > \mu_{\rm min} \approx 313-5$~MeV.
While the pressure vanishes at $\mu_{\rm min}$, 
hadronic (or quark) matter certainly exists, with nonzero pressure, 
for all $\mu > \mu_{\rm min}$.  Thus we imagine that a
very ``soft'' equation of state for massive quarks (and hadrons) matches onto
the equation of state in (\ref{e2}) at some $\mu > \mu_c$ (see also below).  
Consequently, $\mu_c$ cannot be much higher than $\mu_{\rm min}$.

It is this which limits the choice of $\bar\Lambda/\mu$ in
our model.  For $\bar\Lambda = \mu$, $\mu_c = 767$~MeV.
It is absurd to think that the pressure of massive quarks could
be small to densities $\sim 33\rho_0$.  Thus
we do not consider this case further.  

For $\bar\Lambda = 3\mu$, $\mu_c = 300$~MeV when $\alpha_s \sim .6$.
By the Hugenholtz-van Hove theorem, when the pressure
vanishes, the ratio of the total energy 
to the baryon number is $E/A=3\mu_c$.  For iron, $E/A=930$~MeV.  Thus for
$\bar\Lambda = 3\mu$, $E/A = 3 \mu_c = 900$~MeV, and,
as suggested by Bodmer and Witten \cite{bod,wit}, strange
quark matter is absolutely stable relative to hadronic matter.

While possible, we prefer an alternate view.  
Our perturbative equation of state 
is valid only in the chirally symmetric phase, for $\mu > \mu_\chi$.
Perhaps when $\mu < \mu_\chi$, the true equation of state is close
to our perturbative model, but vanishes smoothly as
$\mu \rightarrow \mu_{\rm min}$.  As discussed later,
this is a model for a 
weakly first order (or no) chiral phase transition.  

\begin{figure}[t]
\centerline{\epsfig{figure=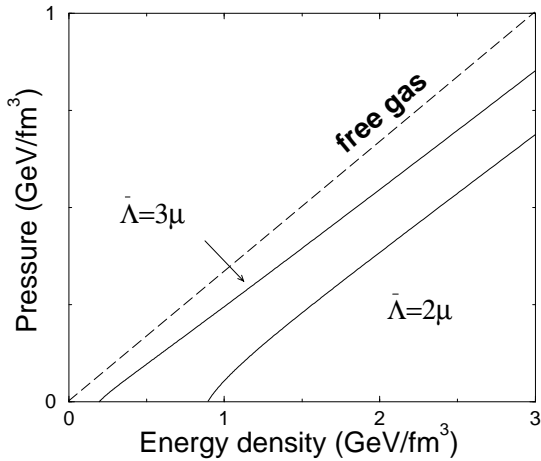,width=7cm}}
\caption{Equation of state for cold quark matter, for
$\bar\Lambda/\mu=2,3$.}
\label{fig:eos}
\end{figure} 

\begin{figure}[t]
\centerline{\epsfig{figure=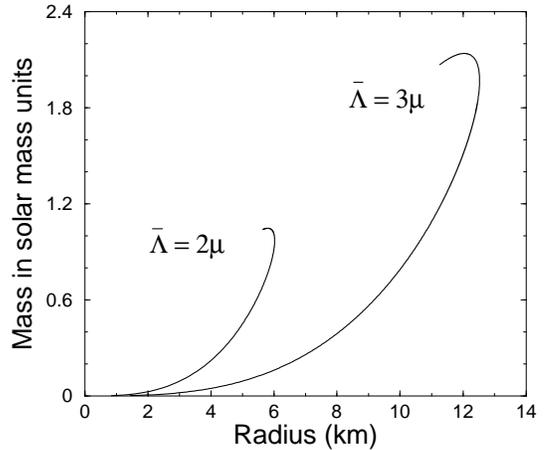,width=7cm}}
\caption{Mass-radius relation of 
the quark star for $\bar\Lambda/\mu=2,3$.}
\label{fig:massradius}
\end{figure}

The structure of a quark star is determined by the solution to the
Tolman-Oppenheimer-Volkov (TOV) equations \cite{G_b}.  
For the TOV equations, all that matters 
is the relationship between pressure and energy density.  This is shown
in Fig.~\ref{fig:eos} for $\bar\Lambda/\mu = 2,3$.  
The numerical solution of the TOV equations 
gives the mass radius relationships of Fig.~\ref{fig:massradius}.  
For any solution to the TOV equation, 
the chemical potential reaches its maximum value at the center of the star;
as one goes out in radius, the chemical potential decreases, and equals
$\mu_c$ (where the pressure vanishes) at the edge of the star.  
For $\bar\Lambda = 3 \mu$, the maximum mass is
$M_{\rm max}\approx 2.14 M_\odot$.
At this mass, the radius is $R_{\rm max}\approx 12$~km; the
chemical potential at the center of the star is $\mu \approx 456$~MeV,
which corresponds to a quark density of 
$\rho_{\rm max} \approx 4.87\rho_0$.
When $\bar\Lambda = 2 \mu$,
$M_{\rm max}\approx 1.05M_\odot$.  At this mass,
the radius is $R_{\rm max}\approx 5.81$~km; the
chemical potential at the center is $\mu \approx 649$~MeV,
corresponding to a quark density of 
$\rho_{\rm max} \approx 14\rho_0$.  

To help understand these results, it is useful to compare to
the equation of state of a nonideal bag model:
\begin{equation}
\Omega(\mu)= - \frac{N_f }{4\pi^2} \; a_{\rm eff} \; \mu^4 + B_{\rm eff} \; ;
\label{e4}
\end{equation}
$B_{\rm eff}$ is an effective bag constant, and the parameter
$a_{\rm eff}$ measures deviations from ideality.
A common choice is to take $a_{\rm eff}$ from the thermodynamic
potential to one loop order, with a fixed value of the coupling constant:
$a_{\rm eff} = 1- 2 \alpha_s/\pi$ \cite{bag}.
In a bag model, the relationship
between pressure and energy density is linear, $p = (\epsilon - 4B)/3$,
{\it irrespective} of the value of $a_{\rm eff}$.  
Thus we can uniformly
scale $p$, $\epsilon$, and $B$ together, so the maximum
mass and radius satisfy a simple scaling relation, 
$M_{\rm max} \sim R_{\rm max} \sim 1/B^{1/2}$ \cite{wit}.

For the chemical potentials of relevance to a quark star,
somewhat surprisingly we find numerically that the pressure in (\ref{e2}) can
be well approximated by the effective bag model, (\ref{e4}).  
This can be seen from Fig.~\ref{fig:eos}, where the relationship
between pressure and energy is very nearly linear.
When $\bar\Lambda = 3\mu$, 
the pressure agrees with a bag model with
$B_{\rm eff}^{1/4} = 157$~MeV and $a_{\rm eff} = .626$ to within
$2\%$ for $\mu: 300 \rightarrow 470$~MeV.  
This is close to the usual value in the MIT bag model,
$B^{1/4} = 145$~MeV \cite{DJJK75}.  When $\bar\Lambda = 2\mu$, the pressure
agrees with a bag model with $B_{\rm eff}^{1/4} = 223$~MeV and
$a_{\rm eff} = .628$ to within $4\%$ for $\mu: 425 \rightarrow 650$~MeV.  

Consequently, the mass radius relationships for our quark stars agree
well with a bag equation of state.  To $\sim 5\%$, the maximum
masses and radii scale according to $\sim 1/B_{\rm eff}^{1/2}$.  
The shape of the mass radius curve is also the same as for
a bag model.  Notably, light quark stars have small radii.
This is because for light stars, $M \ll M_\odot$, the chemical
potential at the center of the star is near $\mu_c$, and the
equation of state is controlled by that of massless fields, minus a bag
constant.

Our results for $\bar\Lambda = 2 \mu$ can be compared to 
other equations of state for dense QCD 
\cite{othera,otherb,thirdb}.
All of these can be viewed as models in which 
there is nonideality at a scale significantly higher than
nuclear matter densities.  Ref. \cite{othera} uses a Schwinger-Dyson
model, and finds 
$M_{\rm max} \approx 0.7M_\odot$ and $R_{\rm max} \approx 9.$~km.
Refs. \cite{otherb} and \cite{thirdb} use models with massive
quasiparticles, so that the masses act as a type of nonideality.
Ref. \cite{otherb} finds
$M_{\rm max} \approx 0.8M_\odot$ and $R_{\rm max} \approx 4.$~km;
ref. \cite{thirdb} finds 
$M_{\rm max} \approx 1.35M_\odot$ and $R_{\rm max} \approx 10.$~km.
We note that in Nambu-Jona-Lasino models, stars with a quark
core do not arise, even at the maximum mass \cite{njl}.

What about the manifestly nonperturbative
phase in which chiral symmetry is spontaneously broken, 
$\mu < \mu_\chi$?
To understand this, consider an expansion in 
a large number of colors \cite{large}.
The usual large $N_c$ limit is to let $N_c \rightarrow \infty$
at fixed $N_f$.  Since quark loops are suppressed in this limit,
however, gluons are only affected by quarks when
$\mu \sim N_c^{1/4}$ or larger.  
This is in contrast to the transition at a nonzero 
temperature, which occurs at a temperature $\sim N_c^0$ \cite{rdp2}.  
We then consider a generalized large $N_c$ limit, in which
$N_f \rightarrow \infty$ at fixed $N_f/N_c$ \cite{general,dum,chiral}.  
The quark thermodynamic potential
is $\Omega \sim - N_f N_c \mu^4$, and the quark
number density $n \sim N_f N_c \mu^3$.  In this limit,
$\mu_\chi \sim N_c^0$, as at nonzero temperature.

For the purposes of our discussion, all that matters is that baryon
masses $m_B \sim N_c$.  The baryon chemical potential is related
to the quark chemical potential as 
$\mu_B = N_c \mu \sim N_c$.  When the baryons
are nonrelativistic, so their Fermi momenta $k_f \sim 1$,
the baryon number density is $n_B \sim d_B k_f^3$.  The degeneracy
of baryons is at least $\sim N_f$, and could easily be larger, 
$\sim N_f^2$.  The baryon thermodynamic potential is (naively)
$\Omega_B \sim - d_B k_f^5/m_B$.  For $k_f \sim 1$, however,
everything is fine: the baryon density is $\sim N_c$ that of
quarks, and $\Omega_B \leq (1/N_c) \Omega $.  In terms of the
quark chemical potential, however, $\mu = \mu_B/N_c$; with
$m_q = m_B/N_c$, $\mu \approx m_q
+ k_f^2/(2 m_B N_c) + \ldots$.  That is, for $k_f \sim 1$,
the region in $\mu$ over which hadrons are a reasonable description
is {\it small}, $\sim 1/N_c^2$.  Also, the binding energy of
nuclear matter is automatically $\sim 1/N_c^2$. 

From this large $N_c$ argument, we conclude that a hadronic description
is applicable {\it only} in a very narrow region of $\mu$; for larger
$\mu$, still of order one, a quark description is appropriate.
This need not be a true phase transition; rather, simply that the 
thermodynamic potential 
may be very difficult to compute in terms of hadrons, but
relatively simple in terms of quarks.  For example,
when $k_f \sim N_c^{1/2}$,
so $\mu - m_q \sim 1/N_c$, naively $\Omega_B \sim - d_B N_c^{3/2}\sim
N_c^{1/2} \Omega$.  This cannot be right --- the thermodynamic
potential of baryons cannot dominate that of quarks.  
The only resolution is that there are cancellations ---
analogous to those which occur for baryon-meson 
couplings \cite{couplings}--- which
greatly reduce the baryon thermodynamic potential, 
so that it is comparable to that
of quarks.  In other words, the hadronic thermodynamic potential {\it must}
``soften'' whenever $\mu - m_q > 1/N_c^2$.

Once one is away from this (narrow) hadronic window in $\mu$, 
the appropriate equation of state for $\mu < \mu_\chi$
is that for massive quarks.  
There are then two possibilities.

The first is
that the thermodynamic potential for massless quarks
matches, more or less smoothly, onto that of massive quarks.
This requires either a weakly first order chiral
transition, or perhaps just crossover.  Below $\mu_\chi$,
as $\mu \rightarrow \mu_{\rm min}$ the quark thermodynamic potential
vanishes in a fashion typical of massive particles; within
$\sim 1/N_c^2$ of $\mu_{\rm min}$, a hadronic description is applicable.

This is illustrated by the choice $\bar\Lambda = 3\mu$.
For a star at its maximum mass, at the center 
$\mu \approx 456$~MeV; as the radius increases, $\mu$ decreases,
until one enters a phase in which chiral symmetry is broken
(assuming $\mu_\chi < 456$~MeV).  Thus
realistically, all of our ``quark'' stars have mantles with
massive quarks, and then hadrons.
For stars near the maximum mass, we assume this mantle is thin,
and does not greatly alter its properties.  As the mass decreases,
however, the portion of the star in the chirally symmetric phase
becomes small, until the entire star is composed entirely of
massive quarks and hadrons.  At this point, the relationship between the stars
mass and radius is no longer like that of 
Fig. \ref{fig:massradius}.  Instead, it looks like that of
nonrelativistic matter, for which the radius increases as the mass
decreases.  It is necessary for the chiral transition to be
weakly first order, or a smooth crossover, for the mass-radius
curve to be continuous.

The second possibility is that the equation of state for massless
quarks does not match smoothly onto that for massive quarks,
with a strongly first order chiral transition \cite{strong}.
As the thermodynamic potential approaches the ideal
gas limit at large $\mu$, and vanishes at $\mu_{\rm min}$, this
requires that the pressure
is small at a value of $\mu_\chi \gg \mu_{min}$.  
Below $\mu_\chi$, by construction 
the pressure of massive quarks is small,
with a ``soft'' equation of state \cite{soft}.  

This occurs if $\bar \Lambda = 2\mu$.  At the maximum mass,
$\mu \sim 649$~MeV at the center; if $\mu_\chi \ll 649$~MeV,
most of the star is composed of massless quarks.
As the mass of the star decreases, so does the amount in the
chirally symmetric phase.  If the chiral phase transition
is strongly first order, eventually one jumps
to a second branch, in which the chemical potentials are always
$< \mu_\chi$ throughout the star.
Stars on this second branch are composed only of massive
quarks and hadrons, with a maximum mass and radius like that of
``ordinary'' neutron stars.
A strong first order chiral transition is necessary to ensure that there 
are two, distinct branches.
Using toy models for the thermodynamic potential, numerically
we obtained solutions to the TOV equations which
display two branches: we patched
a thermodynamic potential 
for massive quarks, for $\mu < \mu_c \approx \mu_\chi$,
onto that for massless quarks, for $\mu > \mu_c$.  In this case,
our stars of massless quarks constitute a third class of compact stars,
after white dwarfs and ``ordinary'' neutron stars \cite{thirda,thirdb}.  

Most pulsars have masses $\sim 1.5 M_\odot$ \cite{G_b}.  
The MACHO project has also reported 
micro-lensing events for the Large Magellanic Cloud
with masses $M=0.15$--0.9$M_\odot$ \cite{MACHO}.
For a weakly first order chiral phase transition, if MACHO events are
hadronic stars, they must have large radii.  For a strongly
first order chiral
phase transition \cite{strong}, MACHO events could be quark stars,
with small radii, and pulsars might represent the second branch.
We stress that our numbers for the maximum mass and radius are
meant only to be suggestive.  Even so, we believe that our
conclusions are qualitatively correct; a weak (or no) chiral phase
transition leads to one type of compact objects, a strongly
first order chiral transition, to two.

 

We thank R. Harlander, J. Lenaghan, A. Peshier, K. Rajagopal, A. Rebhan,
K. Redlich, D. Rischke, and especially L. McLerran for fruitful discussions.
We thank the U.S. Department of Energy for their support under Contract
No. DE-AC02-98CH10886; E.S.F., for the support of CNPq (Brazil);
and J.S.B., for the support of RIKEN and BNL.



\begin{thebibliography}{10}
\bibitem{explanation}
See eq. (4.52) of ref. \cite{Baluni}.  This transformation differs
from that of \cite{Baier}, but agrees with the final
version of \cite{Blaizot_1}.
\bibitem{error}
The uncertainty in $G_0 = 10.374 \pm .13$ is due to errors in numerical
integration.  We take the central value of \cite{Free}; within
errors, it agrees with \cite{Baluni}.
\bibitem{qual}
In (\ref{e3}) we take $\alpha_s(\bar\Lambda)$ to three loop order.
If instead one uses $\alpha_s(\bar\Lambda)$ to two loop order, then
for the same value of $\bar \Lambda/\mu$ ($ = 1,2,3$), 
$\mu_c$ changes by $\sim + 10\%$.  
However, the same $\mu_c$ can be obtained by shifting
$\bar\Lambda/\mu$ by $\sim + 10\%$; with this shift,
the pressure changes by $< 1.5\%$ in all three cases.
\bibitem{soft}
When $\mu < \mu_\chi$, the pressure is small if it 
is dominated by the thermal Wilson line 
(R.~D.~Pisarski, Phys.\ Rev.\ D {\bf 62}, 111501 (2000)),
and if its expectation value is also small.
Why this should be true, 
especially if $N_f \approx N_c$, is unclear \cite{dum}.  Perhaps
$m_s \neq 0$ has a greater effect when $\mu < \mu_\chi$ (kaons
are much heavier than pions) than for $\mu > \mu_\chi$.
\end{thebibliography}
\end{document}